\documentclass[12pt]{article}

\usepackage{amsfonts}			
\newcommand{\RR}{{\mathbb R}}
\newcommand{\CC}{{\mathbb C}}
\newcommand{\HH}{{\mathbb H}}
\newcommand{\OO}{{\mathbb O}}

\usepackage{epsfig}			
\newcommand{\XFig}[3]{\epsfxsize=#3\epsffile[#2]{#1}}

\newcommand{\tr}{{\rm tr\,}}		
\renewcommand{\Re}{{\rm Re}}
\renewcommand{\bar}{\overline}
\renewcommand{\hat}{\widehat}
\renewcommand{\tilde}{\widetilde}
\newcommand{\vf}[1]{\hbox{\boldmath$\vec{#1}$}}
\newcommand{\ii}{\vf\imath}
\newcommand{\jj}{\vf\jmath}
\newcommand{\kk}{\vf k}
\newcommand{\vv}{\vf v}
\newcommand{\ww}{\vf w}
\newcommand\SpinHalf{spin-${1\over2}$}

\newcommand{\Cl}{{\cal C \it l}}
\newcommand{\nxn}{$n \times n$}

\newcommand{\pp}{{\bf p}}
\newcommand{\eplus}{e_{\scriptscriptstyle\uparrow}}
\newcommand{\eminus}{e_{\scriptscriptstyle\downarrow}}
\newcommand{\nupz}{\nu_z}
\newcommand{\numz}{\nu_{{\scriptscriptstyle-}z}}

\begin{document}


\title{\bfseries QUATERNIONIC SPIN
	\thanks{Much of this material was presented in an invited talk
	entitled {\itshape Choosing a Preferred Complex Subalgebra of the
	Octonions} given at the {\bfseries 5th International Conference
	on Clifford Algebras and their Applications in Mathematical Physics}
	in Ixtapa, M\'EXICO, in June 1999.}
}

\author{
	Tevian Dray \\
	{\itshape Department of Mathematics, Oregon State University,
		Corvallis, OR  97331} \\
	{\ttfamily tevian{\rmfamily @}math.orst.edu} \\
		\and
	Corinne A. Manogue \\
	{\itshape Department of Physics, Oregon State University,
		Corvallis, OR  97331} \\
	{\ttfamily corinne{\rmfamily @}physics.orst.edu}
}

\date{1 December 1998 (revised 30 September 1999)}

\maketitle

\begin{abstract}
We rewrite the standard 4-dimensional Dirac equation in terms of quaternionic
\hbox{2-component} spinors, leading to a formalism which treats both massive
and massless particles on an equal footing.  The resulting unified description
has the correct particle spectrum to be a generation of leptons, with the
correct number of spin/helicity states.  Furthermore, precisely three such
generations naturally combine into an octonionic description of the
10-dimensional massless Dirac equation, as previously discussed in~\cite{Dim}.
\end{abstract}

\section{\textbf{INTRODUCTION}}

We recently outlined a new dimensional reduction scheme~\cite{Dim}.  We show
here in detail that applying this mechanism to the 10-dimensional massless
Dirac equation on Majorana-Weyl spinors leads to a quaternionic description of
the full 4-dimensional (free) Dirac equation which treats both massive and
massless particles on an equal footing.  Furthermore, there are naturally 3
such descriptions, each of which corresponds to a generation of leptons with
the correct number of spin/helicity states.

The massive Dirac equation is usually formulated in the context of 4-component
{\it Dirac spinors}.  The 4 degrees of freedom correspond to the choice of
spin (up or down) and the choice of particle or antiparticle.  Similarly,
2-component {\it Penrose spinors}, which can be thought of as the square roots
of null vectors, correspond to massless objects, such as photons.  In
Section~\ref{COMPLEX} we set the stage by reviewing these standard properties
of the chiral description of the momentum-space Dirac equation.

Penrose spinors are usually thought of as Weyl projections of Dirac spinors; a
Dirac spinor contains twice the information of a single Penrose spinor.  As an
alternative to doubling the number of (complex) components, however, we double
the dimension of the underlying division algebra, from the complex numbers
$\CC$ to the quaternions $\HH$.  The anticommutativity of the quaternions then
enables us to package two complex representations of opposite chirality into
the (now quaternionic) 2-component formalism.  In Section~\ref{QUATERNIONIC}
we show how to replace the usual 4-component complex Dirac description with an
equivalent 2-component quaternionic Penrose description, and further discuss
how this puts the massive and massless Dirac equations on an equal footing.

We then consider in Section~\ref{OCTONIONIC} the massless Dirac equation on
Majorana-Weyl spinors (in momentum space) in 10 dimensions, which can be
nicely described in terms of 2-component spinors over the octonions $\OO$, the
only other normed division algebra besides $\RR$, $\CC$, and $\HH$.  Solutions
of this equation are automatically quaternionic, and thus lend themselves to
the preceding quaternionic description.

The final, and most important, ingredient in our approach is the the
dimensional reduction scheme introduced in \cite{Dim}.  In
Sections~\ref{CHOOSING} and~\ref{SPIN} we describe how the choice of a
preferred octonionic unit, or equivalently of a preferred complex subalgebra
$\CC\subset\OO$, naturally reduces 10 spacetime dimensions to 4, and further
allows us to use the standard representation of the Lorentz group $SO(3,1)$ as
$SL(2,\CC)\subset SL(2,\OO)$.  Putting this all together, we show in
Section~\ref{PARTICLES} that the quaternionic spin/helicity eigenstates
correspond precisely to the particle spectrum of a generation of leptons,
consisting of 1 massive and 1 massless particle and their antiparticles.

In Section~\ref{SPINOP}, we discuss the remarkable fact that the quaternionic
spin eigenstates are in fact simultaneous eigenstates of all 3 spin operators,
although the other two eigenvalues are not real.  Finally, in
Section~\ref{DISCUSSION} we discuss our results, in particular that, in a
natural sense, there are precisely 3 such quaternionic subalgebras of the
octonions, which we interpret as generations.

There is a long history of trying to use the quaternions in 4-dimensional
quantum mechanics; see the comprehensive treatment in~\cite{Adler}
and references therein.  Our approach is different in that we use the
additional degrees of freedom to repackage existing information, rather than
increasing the size of the underlying space of scalars.  Ultimately, this
leads us to work in more than 4 spacetime dimensions.

We also note a relatively unknown paper by Dirac~\cite{Dirac} which, much to
our surprise, contains the precursors of several of the key ideas presented
here.

The octonions were first introduced into quantum mechanics by
Jordan~\cite{Jordan,JNW}.  There has in fact been much recent interest in
using the octonions in (higher-dimensional) field theory; excellent modern
treatments can be found in~\cite{GT,Okubo,Dixon}.

After much of this work was completed, we became aware of the recent work of
Sch\"ucking {\it et al.}~\cite{Schucking,SubStandard}, who also use a
quaternionic formalism to describe a single generation of leptons.  They
further speculate that extending the formalism to the octonions would yield a
description of a single generation of quarks as well.  Although the language
is strikingly similar, our approach differs fundamentally from theirs in its
description of momentum.  Ultimately, this hinges on our interpretation of the
obvious $SU(2)$ as spin, whereas Sch\"ucking and coworkers interpret it as
isospin.

\section{\textbf{COMPLEX FORMALISM}}
\label{COMPLEX}

The standard Weyl representation of the gamma matrices in signature
\hbox{($+$ $-$ $-$ $-$)} is
\begin{equation}
\label{CGamma}
\gamma_t = \pmatrix{0& I\cr \noalign{\smallskip} I& 0\cr}
\qquad\qquad
  \gamma_a
  = \pmatrix{0& \sigma_a\cr \noalign{\smallskip} -\sigma_a& 0\cr}
\end{equation}
where $\sigma_a$ for $a=x,y,z$ denote the usual Pauli matrices
\footnote{For later compatibility with our octonion conventions we use $\ell$
rather than $i$ to denote the complex unit.}
\begin{equation}
\sigma_x = \pmatrix{0& 1\cr \noalign{\smallskip} 1& 0\cr} \qquad
  \sigma_y = \pmatrix{0& -\ell\cr \noalign{\smallskip} \ell& 0\cr} \qquad
  \sigma_z = \pmatrix{1& 0\cr \noalign{\smallskip} 0& -1\cr}
\end{equation}
and $I$ is the $2\times2$ identity matrix.

The original formulation of the Dirac equation involves the even part of the
Clifford algebra, historically written in terms of the matrices
$\alpha_a=\gamma_t\gamma_a$ and $\beta=\gamma_t$.  Explicitly, the
momentum-space Dirac equation in this signature can be written as
\begin{equation}
\label{DiracI}
(\gamma_t\gamma_\alpha \, p^\alpha - m \, \gamma_t) \, \Psi = 0
\end{equation}
where $\alpha=t,x,y,z$ and $\Psi$ is a 4-component complex (Dirac) spinor.

Writing $\Psi$ in terms of two 2-component complex Weyl (or Penrose) spinors
$\theta$ and $\eta$ as
\begin{equation}
\Psi = \pmatrix{\theta\cr \noalign{\smallskip} \eta}
\end{equation}
and expanding (\ref{DiracI}) leads to
\begin{equation}
\label{DiracII}
\pmatrix{p^tI-p^a\sigma_a& -m\cr \noalign{\smallskip} -m& p^tI+p^a\sigma_a}
  \pmatrix{\theta\cr \noalign{\smallskip} \eta} = 0
\end{equation}
This leads us to identify the momentum 4-vector with the $2\times2$ Hermitian
matrix
\begin{equation}
\pp
 = p^\alpha \sigma_\alpha
 = \pmatrix{p^t+p^z& p^x-\ell p^y\cr \noalign{\smallskip} p^x+\ell p^y& p^t-p^z}
\end{equation}
where we have set $\sigma_t=I$, which reduces (\ref{DiracII}) to the two
equations
\begin{eqnarray}
  -\tilde\pp\,\theta - m\eta &=& 0 \label{DiracIIIa}\\
  -m\theta + \pp\,\eta &=& 0 \label{DiracIIIb}
\end{eqnarray}
where the tilde denotes trace-reversal.  Explicitly,
\begin{equation}
\tilde\pp = \pp - \tr(\pp) I
\end{equation}
which reverses the sign of $p^t$, so that $-\tilde\pp$ can be identified with
the 1-form dual to $\pp$.  This interpretation is strengthened by noting that
\begin{equation}
-\tilde\pp \pp = \det(\pp) I = p_\alpha p^\alpha I = m^2 I
\end{equation}
where the identification of the norm of $p^\alpha$ with $m$ is just the
compatibility condition between (\ref{DiracIIIa}) and (\ref{DiracIIIb}).

\section{\textbf{QUATERNIONIC FORMALISM}}
\label{QUATERNIONIC}

The {\it quaternions} $\HH$ are the associative, noncommutative, normed
division algebra over the reals.  The quaternions are spanned by the identity
element $1$ and three imaginary units, usually denoted $i$, $j$, $k:=ij$.
Quaternionic conjugation, denoted with a bar, is given by reversing the sign
of each imaginary unit.  Each imaginary unit squares to $-1$, and they
anticommute with each other; the full multiplication table then follows using
associativity.%
\footnote{The use of $\ii$, $\jj$, $\kk$ for Cartesian basis vectors
originates with the quaternions, which were introduced by Hamilton as an early
step towards vectors~\cite{Crowe}.  Making the obvious identification of
vectors $\vv$, $\ww$ with imaginary quaternions $v$, $w$, then the real part
of the quaternionic product $vw$ is (minus) the dot product $\vv\cdot\ww$,
while the imaginary part is the cross product $\vv\times\ww$.}  However, in
order to avoid conflict with our subsequent conventions for the octonions, we
will instead label our quaternionic basis $\ell$, $k$, $\ell k$.  The
imaginary unit $\ell$ will play the role of the complex unit $i$, and, as we
will see later, $k$ will label this particular quaternionic subalgebra of the
octonions.  In terms of the Cayley-Dickson process~\cite{Dickson,Schafer}, we
have
\begin{equation}
\label{Cayley}
\HH = \CC + \CC k = (\RR + \RR\ell) + (\RR + \RR\ell)k
\end{equation}

As vector spaces, $\HH = \CC^2$, which allows us to identify $\HH^2$ with
$\CC^4$ in several different ways.  We choose the identification
\begin{equation}
\label{IdentifyC}
\pmatrix{A\cr B\cr C\cr D\cr}
  \longleftrightarrow
  \pmatrix{C - k B\cr \noalign{\smallskip} D + k A\cr}
\end{equation}
with $A,B,C,D\in\CC$.  Equivalently, we can write this identification in terms
of the Weyl (Penrose) spinors $\theta$ and $\eta$ as
\begin{equation}
\label{IdentifyK}
\Psi = 
\pmatrix{\theta\cr \noalign{\smallskip} \eta\cr}
  \longleftrightarrow
  \eta + \sigma_k \theta
\end{equation}
where we have introduced the generalized Pauli matrix
\begin{equation}
\sigma_k = \pmatrix{0& -k\cr \noalign{\smallskip} k& 0\cr}
\end{equation}

Since (\ref{IdentifyK}) is clearly a vector space isomorphism, there is also
an isomorphism relating the linear maps on these spaces.  We can use the
induced isomorphism to rewrite the Dirac equation (\ref{DiracI}) in
2-component quaternionic language.  Direct computation yields the
correspondences
\begin{equation}
\gamma_t \gamma_a 
  = \pmatrix{-\sigma_a& 0\cr \noalign{\smallskip} 0& \sigma_a\cr}
  \longleftrightarrow
  \sigma_a
\end{equation}
and
\begin{equation}
\gamma_t \longleftrightarrow \sigma_k
\end{equation}
and of course also
\begin{equation}
\gamma_t\gamma_t \longleftrightarrow \sigma_t
\end{equation}
since the left-hand-side is the $4\times4$ identity matrix and the
right-hand-side is the $2\times2$ identity matrix.  Direct translation of
(\ref{DiracI}) now leads to the quaternionic Dirac equation
\begin{equation}
\label{DiracIV}
(\pp - m\sigma_k) (\eta + \sigma_k\theta) = 0
\end{equation}
Working backwards, we can separate this into an equation not involving $k$,
which is precisely (\ref{DiracIIIb}), and an equation involving $k$, which is
\begin{equation}
\label{DiracV}
\pp \, \sigma_k \theta - m \sigma_k \eta = 0
\end{equation}
Multiplying this equation on the left by $\sigma_k$, and using the remarkable
identity
\begin{equation}
\label{Remarkable}
\sigma_k \, \pp \, \sigma_k = - \tilde\pp
\end{equation}
reduces (\ref{DiracV}) to (\ref{DiracIIIa}), as expected.

So far, we have done nothing more or less than rewrite the usual
momentum-space Dirac equation in 2-component quaternionic language.  However,
the appearance of the term $m\sigma_k$ suggests a way to put in the mass term
on the same footing as the other terms, which we now exploit.
Multiplying (\ref{DiracIV}) on the left by $-\sigma_k$ and using
(\ref{Remarkable}) brings the Dirac equation to the form
\begin{equation}
\label{DiracVI}
(\tilde\pp+m\sigma_k) \, \psi = 0
\end{equation}
where we have introduced the 2-component quaternionic spinor
\begin{equation}
\psi = \sigma_k (\eta + \sigma_k\theta) = \theta + \sigma_k\eta
\end{equation}
When written out in full, (\ref{DiracVI}) takes the form
\begin{equation}
\pmatrix{-p^t+p^z& p^x-\ell p^y-km\cr
	   \noalign{\smallskip}
	   p^x+\ell p^y+km& -p^t-p^z}
  \psi = 0
\end{equation}
This clearly suggests viewing the mass as an additional spacelike component of
a higher-dimensional vector.  Furthermore, since the matrix multiplying $\psi$
has determinant zero, this higher-dimensional vector is null.  We thus appear
to have reduced the massive Dirac equation in 4 dimensions to the massless
Dirac, or Weyl, equation in higher dimensions, thus putting the massive and
massless cases on an equal footing.  This expectation is indeed correct, as we
show in the next section in the more general octonionic setting.

\section{\textbf{OCTONIONIC FORMALISM}}
\label{OCTONIONIC}

\subsection{Octonionic Penrose Spinors}

\begin{figure}
\begin{center}
\XFig{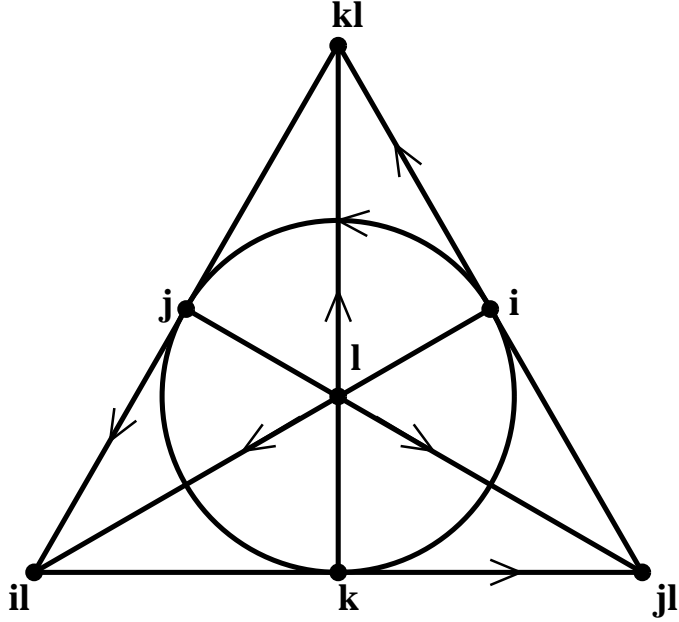}{68 168 543 614}{3.5in}
\end{center}
\caption{The representation of the octonionic multiplication table using the
7-point projective plane.  Each of the 7 oriented lines gives a
quaternionic triple.}
\label{Omult}
\end{figure}

The {\it octonions} $\OO$ are the nonassociative, noncommutative, normed
division algebra over the reals.  The octonions are spanned by the identity
element $1$ and seven imaginary units, which we label as
$\{i,j,k,k\ell,j\ell,i\ell,\ell\}$.  Each imaginary unit squares to $-1$
\begin{equation}
i^2 = j^2 = k^2 = ... = \ell^2 = -1
\end{equation}
and the full multiplication table can be conveniently encoded in the 7-point
projective plane, as shown in Figure~\ref{Omult}; each line is to be thought of
as a circle.  The octonionic units can be grouped into (the imaginary parts
of) quaternionic subalgebras in 7 different ways, corresponding to the 7 lines
in the figure; these will be referred to as quaternionic triples.  Within each
triple, the arrows give the orientation, so that e.g.
\begin{equation}
ij = k = -ji
\end{equation}
Any three imaginary basis units which do not lie in a such a triple
anti-associate.  Note that any two octonions automatically lie in (at least
one) quaternionic triple, so that expressions containing only two independent
imaginary octonionic directions do associate.  {\it Octonionic conjugation} is
given by reversing the sign of the imaginary basis units, and the norm is just
\begin{equation}
|p| = \sqrt{p \bar{p}}
\end{equation}
which satisfies the defining property of a normed division algebra, namely
\begin{equation}
|pq| = |p| |q|
\end{equation}

We follow~\cite{SUDBERY,Chung} in representing real $(9+1)$-dimensional
Minkowski space in terms of $2\times2$ Hermitian octonionic matrices.
\footnote{A number of authors, such as~\cite{KT}, have used this approach to
describe supersymmetric theories in 10 dimensions.  Fairlie \&
Manogue~\cite{FairlieI,FairlieII} and Manogue \& Sudbery~\cite{Sudbery}
described solutions of the superstring equations of motion using octonionic
parameters, and Schray~\cite{Schray,Thesis} described the superparticle.  A
more extensive bibliography appears in~\cite{Schray}.}
In analogy with the complex case, a vector field $q^\mu$ with $\mu=0,...,9$
can be thought of under this representation as a matrix
\begin{equation}
Q = \pmatrix{~~q^+& \bar{q}\cr \noalign{\smallskip} q& ~~q^-}
\end{equation}
where $q^\pm=q^0\pm q^9 \in\RR$ are the components of $q^\mu$ in 2 null
directions and $q=q^1 +q^2 i + ... + q^8 \ell \in\OO$ is an octonion
representing the transverse spatial coordinates.  Following \cite{Sudbery}, we
define
\begin{equation}
\tilde{Q} = Q - \tr(Q) I
\end{equation}
Furthermore, since $Q$ satisfies its characteristic polynomial, we have
\begin{equation}
-Q\tilde{Q} = -\tilde{Q}Q = -Q^2 + \tr(Q) Q = \det(Q) I
  = g_{\mu\nu} q^\mu q^\nu I
\end{equation}
where $g_{\mu\nu}$ is the Minkowski metric (with signature ($+$ $-$ ... $-$)).
We can therefore identify the tilde operation with the metric dual, so that
$-\tilde{Q}$ represents the covariant vector field $q_\mu$.

Just as in the complex case (compare~(\ref{CGamma})), this can be thought of
(up to associativity issues) as a Weyl representation of the underlying
Clifford algebra $\Cl(9,1)$ in terms of $4\times4$ gamma matrices of the form
\begin{equation}
\label{MatDef}
q^\mu \gamma_\mu
  = \pmatrix{0&Q\cr \noalign{\smallskip} -\tilde{Q}&0\cr}
\end{equation}
which are now octonionic.  It is readily checked that
\begin{equation}
\gamma_\mu\gamma_\nu + \gamma_\nu\gamma_\mu = 2\,g_{\mu\nu}
\end{equation}
as desired.

In this language, a Majorana spinor
$\Psi=\pmatrix{\psi\cr\noalign{\smallskip}\chi}$ is a 4-component octonionic
column, whose chiral projections are the Majorana-Weyl spinors
$\pmatrix{\psi\cr\noalign{\smallskip}0}$ and
$\pmatrix{0\cr\noalign{\smallskip}\chi}$, which can be identified with the
2-component octonionic columns $\psi$ and $\chi$, which in turn can be thought
of as generalized Penrose spinors.  Writing
\begin{equation}
\gamma_\mu
  = \pmatrix{0& \sigma_\mu\cr \noalign{\smallskip}-\tilde\sigma_\mu& 0\cr}
\end{equation}
or equivalently
\begin{equation}
\label{Qdef}
Q = q^\mu \sigma_\mu = q_\mu \sigma^\mu
\end{equation}
defines the {\it octonionic Pauli matrices} $\sigma_\mu$.  The matrices
$\sigma_a$, with $a=1,...,9$, are the natural generalization of the ordinary
Pauli matrices to the octonions, and $\sigma_0=I$.  In analogy with our
treatment of the complex case, we have
\begin{equation}
\label{OctoI}
\gamma_0 \gamma_\mu
    = \pmatrix{-\tilde\sigma_\mu& 0\cr \noalign{\smallskip} 0& \sigma_\mu\cr}
\end{equation}

For completeness, we record some useful relationships.
The adjoint $\bar\Psi$ of the Majorana spinor $\Psi$ is given as usual by
\begin{equation}
\bar\Psi = \Psi^\dagger \gamma_0
\end{equation}
since
\begin{equation}
\gamma_\mu^\dagger \gamma_0^\dagger = \gamma_0 \gamma_\mu
\end{equation}
Given a Majorana spinor $\Psi=\pmatrix{\psi\cr\noalign{\smallskip}\chi}$, we
can construct a real vector
\footnote{We assume here that the components of our spinors are {\it
commuting}, as we believe that the anticommuting nature of fermions may be
carried by the octonionic units themselves.  An analogous result for
anticommuting spinors was obtained in both of~\cite{FairlieII,Schray}.}
\begin{equation}
q^\mu[\Psi] = \Re(\Psi^\dagger\gamma^0\gamma^\mu\Psi)
\end{equation}
corresponding in traditional language to $\bar\Psi\gamma^\mu\Psi$.  We can
further identify this with a $2\times2$ matrix $Q[\Psi]$ as in (\ref{MatDef})
above.  Direct computation using the cyclic property of the trace, e.g.\ for
octonionic columns $\Psi_1$, $\Psi_2$, and octonionic matrices $\gamma$
\begin{equation}
\Re(\Psi_1^\dagger \gamma \Psi_2^{\phantom\dagger})
  = \Re\left(\tr(\Psi_1^\dagger \gamma \Psi_2^{\phantom\dagger})\right)
  = \Re\left(\tr(\Psi_2^{\phantom\dagger} \Psi_1^\dagger \gamma)\right)
\end{equation}
shows that~\cite{FairlieI,FairlieII}
\begin{equation}
\label{MSq}
Q[\Psi] = 2 \, \psi\psi^\dagger - 2 \, \tilde{\chi\chi^\dagger}
\end{equation}

\subsection{Octonionic Dirac Equation}

The momentum-space massless Dirac equation (Weyl equation) in 10 dimensions
can be written in the form
\begin{equation}
\label{Weyl}
\gamma_0\gamma_\mu \, p^\mu \, \Psi = 0
\end{equation}
Choosing $\Psi=\pmatrix{\psi\cr\noalign{\smallskip}0}$ to be a Majorana-Weyl
spinor, and using (\ref{OctoI}) and (\ref{Qdef}), (\ref{Weyl}) finally takes
the form
\begin{equation}
\label{StringII}
\tilde{P}\psi = 0
\end{equation}
which is the octonionic Weyl equation.  In matrix notation, it is
straightforward to show that the momentum $p^\mu$ of a solution of the Weyl
equation must be null: (\ref{StringII}) says that the $2\times2$ Hermitian
matrix $P$ has $0$ as one of its eigenvalues,
\footnote{It is {\it not} true in general~\cite{Eigen,Find} that the
determinant of an \nxn\ Hermitian octonionic matrix is the product of its
(real) eigenvalues, unless $n=2$; however, see also~\cite{Other}.}
which forces $\det(P)=0$, which is
\begin{equation}
\label{StringI}
\tilde{P}P = 0
\end{equation}
which in turn is precisely the condition that $p^\mu$ be null.

Equations (\ref{StringI}) and (\ref{StringII}) are algebraically the same as
the octonionic versions of two of the superstring equations of motion, as
discussed in~\cite{FairlieI,FairlieII,Sudbery}, and are also the octonionic
superparticle equations~\cite{Schray}.  As implied by those references,
(\ref{StringI}) implies the existence of a 2-component spinor $\theta$ such
that
\begin{equation}
\label{SolI}
P = \pm\theta\theta^\dagger
\end{equation}
where the sign corresponds to the time orientation of $P$, and the general
solution of (\ref{StringII}) is
\begin{equation}
\label{SolII}
\psi = \theta\xi
\end{equation}
where $\xi\in\OO$ is arbitrary.  The components of $\theta$ lie in the complex
subalgebra of $\OO$ determined by $P$, so that (the components of) $\theta$
and $\xi$ (and hence also $P$) belong to a quaternionic subalgebra of $\OO$.
Thus, for solutions (\ref{SolII}), the Weyl equation (\ref{Weyl}) itself
becomes quaternionic.

Furthermore, it follows immediately from (\ref{SolII}) that
\begin{equation}
\label{Proportional}
\psi\psi^\dagger = \pm |\xi|^2 P
\end{equation}
Comparing this with (\ref{MSq}), we see that the vector constructed from
$\psi$ is proportional to $P$, or in more traditional language
\begin{equation}
\bar\Psi\gamma^\mu\Psi \sim p^\mu
\end{equation}
which can be interpreted as the requirement that the Pauli-Lubanski spin
vector be proportional to the momentum for a massless particle.

\section{\textbf{DIMENSIONAL REDUCTION AND SPIN}}

\subsection{Choosing a Preferred Complex Subalgebra}
\label{CHOOSING}

The description in the preceding section of 10-dimensional Minkowski space in
terms of Hermitian octonionic matrices is a direct generalization of the usual
description of ordinary (4-dimensional) Minkowski space in terms of complex
Hermitian matrices.  If we fix a complex subalgebra $\CC\subset\OO$, then we
single out a 4-dimensional Minkowski subspace of 10-dimensional Minkowski
space.  The projection of a 10-dimensional null vector onto this subspace is a
causal 4-dimensional vector, which is null if and only if the original vector
was already contained in the subspace, and timelike otherwise.  The time
orientation of the projected vector is the same as that of the original, and
the induced mass is given by the norm of the remaining 6 components.
Furthermore, the ordinary Lorentz group $SO(3,1)$ clearly sits inside the
Lorentz group $SO(9,1)$ via the identification of their double-covers, the
spin groups $\hbox{Spin}(d,1)$, namely
\footnote{The last equality is more usually discussed at the Lie algebra
level.  Manogue \& Schray~\cite{Lorentz} gave an explicit representation using
this language of the {\it finite} Lorentz transformations in 10 spacetime
dimensions.  For further discussion of the notation $SL(2,\OO)$, see
also~\cite{Mobius}.}
\begin{equation}
\hbox{Spin}(3,1) = SL(2,\CC) \subset SL(2,\OO) = \hbox{Spin}(9,1)
\end{equation}

Therefore, all it takes to break 10 spacetime dimensions to 4 is to choose a
preferred octonionic unit to play the role of the complex unit.  We choose
$\ell$ rather than $i$ to fill this role, preferring to save $i$, $j$, $k$ for
a (distinguished) quaternionic triple.  The projection $\pi$ from $\OO$ to
$\CC$ is given by
\begin{equation}
\pi(q) = {1\over2} (q + \ell q \bar\ell)
\end{equation}
and we thus obtain a preferred $SL(2,\CC)$ subgroup of $SL(2,\OO)$,
corresponding to the ``physical'' Lorentz group.

\subsection{Spin}
\label{SPIN}

Since we now have a preferred 4-d Lorentz group, we can use its rotation
subgroup $SU(2)\subset SL(2,\CC)$ to define spin.  However, care must be taken
when constructing the Lie algebra $su(2)$, due to the lack of commutativity.

Under the usual action of $M\in SU(2)$ on a Hermitian matrix $Q$ (thought of
as a spacetime vector via (\ref{Qdef})), namely
\begin{equation}
Q \mapsto MQM^\dagger
\end{equation}
we can identify the basis rotations as usual as
\begin{equation}
R_z = \pmatrix{e^{\ell{\phi\over2}}& 0\cr
		\noalign{\smallskip} 0&
		e^{-\ell{\phi\over2}}\cr}
\qquad
R_y = \pmatrix{~~\cos{\phi\over2}& \sin{\phi\over2}\cr
		\noalign{\smallskip}
		-\sin{\phi\over2}& \cos{\phi\over2}\cr}
\qquad
R_x = \pmatrix{\cos{\phi\over2}& \ell\sin{\phi\over2}\cr
		\noalign{\smallskip}
		\ell\sin{\phi\over2}& \cos{\phi\over2}\cr}
\end{equation}
corresponding to rotations by the angle $\phi$ about the $z$, $y$, and $x$ axes,
respectively.

The infinitesimal generators of the Lie algebra $su(2)$ are obtained by
differentiating these group elements, via
\begin{equation}
L_a = {d R_a\over d\phi} \Bigg|_{\phi=0}
\end{equation}
where as before $a=x,y,z$.  For reasons which will become apparent, we have
{\it not} multiplied these generators by $-\ell$ to obtain Hermitian matrices.
We have instead
\begin{equation}
2 L_z = \pmatrix{\ell& 0\cr \noalign{\smallskip} 0& -\ell\cr}
\qquad
2 L_y = \pmatrix{0& 1\cr \noalign{\smallskip} -1& 0\cr}
\qquad
2 L_x = \pmatrix{0& \ell\cr \noalign{\smallskip} \ell& 0\cr}
\end{equation}
which satisfy the commutation relations
\begin{equation}
\left[ L_a , L_b \right] = \epsilon_{abc} L_c
\end{equation}
where $\epsilon$ is completely antisymmetric and
\begin{equation}
\epsilon_{xyz} = 1
\end{equation}

Spin eigenstates are usually obtained as eigenvectors of the Hermitian matrix
$-\ell L_z$, with real eigenvalues.  Here we must be careful to multiply by
$\ell$ in the correct place.  We define
\begin{equation}
\hat{L}_z \psi := - L_z \psi \ell
\end{equation}
which is well-defined by alternativity, so that
\begin{equation}
\hat{L}_z = - \ell_R \circ L_z
\end{equation}
where the operator $\ell_R$ denotes right multiplication by $\ell$ and where
$\circ$ denotes composition.  The operators $\hat{L}_a$ are self-adjoint with
respect to the inner product
\begin{equation}
\label{SelfAdj}
\langle \psi,\chi \rangle = \pi \!\left( \psi^\dagger\chi \right)
\end{equation}
We therefore consider the eigenvalue problem
\begin{equation}
\label{HEigen}
\hat{L}_z \psi = \psi \lambda
\end{equation}
with $\lambda\in\RR$.  It is straightforward to show that the real eigenvalues
are
\begin{equation}
\lambda_\pm = \pm {1\over2}
\end{equation}
as expected.  However, the form of the eigenvectors is a bit more surprising:
\begin{equation}
\psi_+ = \pmatrix{A\cr kD\cr} \qquad\qquad
  \psi_- = \pmatrix{kB\cr C\cr}
\end{equation}
where $A,B,C,D\in\CC$ are any elements of the preferred complex subalgebra,
and $k$ is any imaginary octonionic unit orthogonal to $\ell$, so that $k$ and
$\ell$ anticommute.  Thus, the components of spin eigenstates are contained in
the quaternionic subalgebra $\HH\subset\OO$ which is generated by $\ell$ and
$k$.

Therefore, if we wish to consider spin eigenstates, $\ell$ must be in the
quaternionic subalgebra $\HH$ defined by the solution.  We can further assume
without loss of generality that $\HH$ takes the form given in~(\ref{Cayley}).
Thus, the only possible nonzero components of $p_\mu$ are $p_t=p_0$,
$p_x=p_1$, $p_k=p_4$, $p_{k\ell}=p_5$, $p_y=p_8$, and $p_z=p_9$, corresponding
to the gamma matrices with components in $\HH$.  We can further assume (via a
rotation in the ($k,k\ell$)-plane if necessary) that $p_5=0$, so that
\begin{equation}
\label{PComplex}
P = \pi(P) + m \, \sigma^k
\end{equation}
where
\begin{equation}
\pi(P) = p_\alpha \sigma^\alpha \equiv \pp
\end{equation}
with $\alpha = 0,1,8,9$ (or equivalently $\alpha=t,x,y,z$) is complex, and
corresponds to the 4-dimensional momentum of the particle, with squared mass
\begin{equation}
m^2 = p_\alpha p^\alpha = -\det(\pi(P))
\end{equation}
Inserting (\ref{PComplex}) into (\ref{StringII}), we recover precisely
(\ref{DiracVI}), and we see that we have come full circle: Solutions of the
{\it octonionic\/} Weyl equation~(\ref{Weyl}) are described precisely by the
{\it quaternionic\/} formalism of Section~\ref{QUATERNIONIC}, and the
dimensional reduction scheme determines the mass term.

\subsection{Particles}
\label{PARTICLES}

For each solution $\psi$ of (\ref{SolII}), the momentum is proportional to
$\psi\psi^\dagger$ by (\ref{Proportional}).  Up to an overall factor, we can
therefore read off the components of the 4-dimensional momentum $p_\alpha$
directly from $\pi(\psi\psi^\dagger)$.  We can use a Lorentz transformation to
bring a massive particle to rest, or to orient the momentum of a massless
particle to be in the $z$-direction.

If $m\ne0$, we can distinguish particles from antiparticles by the sign of the
term involving $m$, which is the coefficient of $\sigma_k$ in $P$.
Equivalently, we have the particle/antiparticle projections (at rest)
\begin{equation}
\Pi_\pm
  = {1\over2} \left( \sigma_t \pm \sigma_k \right)
\end{equation}
If $m=0$, however, we can only distinguish particles from antiparticles in
momentum space by the sign of $p^0$, as usual; this is the same as the sign in
(\ref{Proportional}).  Similarly, in this language, the chiral projection
operator is constructed from
\begin{equation}
\Upsilon^5
  = \sigma^t \sigma^x \sigma^y \sigma^z
  = - \pmatrix{\ell& 0\cr 0& \ell\cr}
\end{equation}
However, as with spin, we must multiply by $\ell$ in the correct place,
obtaining
\begin{equation}
\hat\Upsilon^5 = \ell_R \circ \Upsilon^5
\end{equation}
As a result, even though $\Upsilon^5$ is a multiple of the identity,
$\hat\Upsilon^5$ is not, and the operators
${1\over2}(\sigma_t\pm\hat\Upsilon^5)$ project $\HH^2$ into the Weyl subspaces
$\CC^2\oplus\CC^2 k$ as desired.

Combining the spin and particle information, over the quaternionic subalgebra
$\HH\subset\OO$ determined by $k$ and $\ell$, we thus find 1 massive
\SpinHalf\ particle at rest, with 2 spin states, namely
\begin{equation}
\eplus = \pmatrix{1\cr k\cr}
\qquad
  \eminus = \pmatrix{-k\cr ~~1\cr}
\end{equation}
whose antiparticle is obtained by replacing $k$ by $-k$ (and changing the sign
in (\ref{Proportional})).  We also find 1 massless \SpinHalf\ particle
involving $k$ moving in the $z$-direction, with a single helicity state,
\begin{equation}
\nupz = \pmatrix{0\cr k\cr}
\end{equation}
which corresponds, as usual, to both a particle and its antiparticle.  It is
important to note that
\begin{equation}
\numz = \pmatrix{k\cr 0\cr}
\end{equation}
corresponds to a massless particle with the same helicity moving in the
opposite direction, not to a different particle with the opposite helicity.
Each of the above states may be multiplied (on the {\it right}) by an
arbitrary complex number.

There is also a single {\it complex\/} massless \SpinHalf\ particle, with the
opposite helicity, which is given in momentum space by
\begin{equation}
\label{Oz}
\hbox{\O}_z=\pmatrix{0\cr 1\cr}
\end{equation}
As with the other massless momentum space states, this describes both a
particle and an antiparticle.  Alone among the particles, this one does not
contain $k$, and hence does not depend on the choice of identification of a
particular quaternionic subalgebra $\HH$ satisfying $\CC\subset\HH\subset\OO$.

\subsection{Spin Operators}
\label{SPINOP}

We saw in the previous section that the spin up particle state $\eplus$ is a
simultaneous eigenvector of the spin operator and particle projections, that
is
\begin{equation}
2\hat{L}_z \eplus = \eplus = \Pi_+ \eplus
\end{equation}
Remarkably, $\eplus$ is also an eigenvector of the remaining spin operators,
namely
\begin{equation}
2\hat{L}_x \eplus = -\eplus \,k \qquad
  2\hat{L}_y \eplus = -\eplus \,k\ell
\end{equation}
although the eigenvalues are not real.  Similar statements hold for the
corresponding spin down and antiparticle states, although with different
eigenvalues.

We find it illuminating to consider the equivalent right eigenvalue
problem
\begin{equation}
L_z \psi = \psi \lambda
\end{equation}
for the non-Hermitian operator $L_z$.  The operator $L_z$ admits imaginary
eigenvalues $\pm \ell/2$, which correspond to the usual spin eigenstates.  But
$L_z$ also admits other imaginary eigenvalues!  These correspond precisely to
the eigenvalues of $\hat{L}_z$ which are not real, and in fact not in $\CC$.
We emphasize that the spin operators {\it are\/} self-adjoint (with respect
to~(\ref{SelfAdj})).  However, over the octonions it is not true that all the
eigenvalues of Hermitian matrices, are real~\cite{Eigen,NonReal}; the case of
self-adjoint operators is similar.

How does it affect the traditional interpretation of quantum mechanics to have
simultaneous eigenstates of all 3 spin operators?  The essential feature
which permits this is that only one of the eigenvalues is real, and only real
eigenvalues correspond to observables.  Thus, from this point of view, the
reason that the spin operators fail to commute is not that they do not admit
simultaneous eigenstates, but rather that their {\it eigenvalues\/} fail to
commute!

Furthermore, while (right) multiplication by a (complex) phase does not change
any of the real eigenvalues, the nonreal eigenvalues do depend on the phase,
since the phase doesn't commute with the eigenvalue!  Does this allow one
{\it in principle\/} to determine the exact spin orientation, even if no
corresponding measurement exists?

\section{\textbf{DISCUSSION}}
\label{DISCUSSION}

We have shown how the massless Dirac equation in 10 dimensions reduces to the
(massive and massless) Dirac equation in 4 dimensions when a preferred
octonionic unit is chosen.

The quaternionic Dirac equation discussed in Section~\ref{QUATERNIONIC}
describes 1 massive particle with 2 spin states, 1 massless particle with only
1 helicity, and their antiparticles.  We identify this set of particles with a
generation of leptons.

Furthermore, as can be seen from Figure~\ref{Omult}, there is room in the
octonions for exactly 3 such quaternionic descriptions which have only their
complex part in common, corresponding to replacing $k$ in turn by $i$ and $j$.
We identify these 3 quaternionic spaces as describing 3 generations of
leptons.

There is, however, one additional massless particle/antiparticle pair, given
by (\ref{Oz}).  Being purely complex, it does not belong to any generation,
and it has the opposite helicity from the other massless particles.  We do not
currently have a physical interpretation for this additional particle; if this
theory is to correspond to nature, then this additional particle must for some
reason not interact much with anything else.

Note that the mass appears in this theory as an overall scale, which can be
thought of as the length scale associated with the corresponding quaternionic
direction.  In particular, antiparticles must have the same mass as the
corresponding particles.  This suggests that the only free parameters in this
theory are 3 length scales, corresponding to the masses in each generation.

The theory presented here can be elegantly rewritten in terms of {\it Jordan
matrices}, i.e.\ $3\times3$ octonionic Hermitian matrices, along the lines of
the approach to the superparticle presented in~\cite{Schray}.  This approach,
which is briefly described in~\cite{Other}, demonstrates that the theory is
invariant under a much bigger group than the Lorentz group, namely the
exceptional group $E_6$ (actually, $E_7$, since only conformal transformations
are involved).  We therefore believe it may be possible to extend the theory
so as to include quarks and color.

Finally, as noted in~\cite{Dim}, we have worked only in momentum space, and
have discussed only free particles.  Perhaps our most intriguing result is the
observation that the introduction of position space would require a preferred
complex unit in the Fourier transform.  Similarly, a description of
interactions based on minimal coupling would again involve a preferred complex
unit.  Therefore, it does not appear to be {\it possible} to use the formalism
presented here to give a full, interacting, 10-dimensional theory in which all
10 spacetime dimensions are on an equal footing.  We view this as a
tantalizing hint that not only interactions, but even 4-dimensional spacetime
itself, may arise as a consequence of the symmetry breaking from 10 dimensions
to 4!

\bigskip\leftline{\bf ACKNOWLEDGMENTS}\nobreak

It is a pleasure to thank David Griffiths, Phil Siemens, Tony Sudbery, and Pat
Welch for comments on earlier versions of this work, Paul Davies for moral
support, and Reed College for hospitality.

\end{document}